\documentclass[pra,aps,amssymb,twocolumn,noshowpacs,hyperref]{revtex4-1}
\usepackage{color}
\usepackage{graphicx}
\usepackage{hyperref}
\usepackage{amsmath}
\usepackage{latexsym}
\usepackage{amssymb}
\usepackage{mathrsfs}
\usepackage{mathtools}
\usepackage{layout}
\usepackage{verbatim}
\usepackage{multirow}
\usepackage{bm}
\usepackage{amsfonts,epsfig}
\usepackage{lineno}
\usepackage{textcomp}
\usepackage{diagbox}

\begin{document}
\title{High-fidelity molecular quantum logic gates resilient to interaction fluctuation}
\author{Yan Lu}
\affiliation{Center for Theoretical Physics and School of Physics and Optoelectronic Engineering, Hainan University, Haikou 570228, China}
\author{Xiao-Feng Shi}
\affiliation{Center for Theoretical Physics and School of Physics and Optoelectronic Engineering, Hainan University, Haikou 570228, China}
\begin{abstract}
Optically trapped polar molecules are promising for quantum information processing, yet the accuracy of an entangling molecular gate is limited by the uncertainty of dipole-dipole interactions~(DDI) from the molecular motion in traps. We show that two $\pi$ pulses of global microwave excitation can yield a high-fidelity controlled-phase gate when assisted by two single-qubit gates. The gate is resilient to the uncertainty of DDI because it does not rely on populating DDI-coupled states. Further, the controlled phase is fully tunable by varying the relative phase of the two global microwave pulses, and, hence, the gate can find applications in a wide range of quantum algorithms involving quantum Fourier transform. Moreover, we introduce a motional-mode separation technique to quantum mechanically study the influence of the molecular motion, which shows that the gate fidelity can be over 0.9999 with typical experimental conditions.
\end{abstract}
\maketitle

\section{introduction}
Polar molecules were proposed as candidates for building quantum information processing devices more than two decades ago~\cite{DeMille_2002}, with multiple follow-up designs for molecular quantum entangling gates~\cite{Zhu_2013,Yelin_2006,Ni_2018,Hughes_2020,Tscherbul_2023,bergonzoni_iswap_2025,Muminov_2026}. These gate protocols depend on shelving molecules transiently on a state coupled by dipole-dipole interaction~(DDI), which can be either a state-dependent Ising-like energy shift~\cite{Yelin_2006,Tscherbul_2023}, or a spin-exchange coupling~\cite{Ni_2018,Hughes_2020,bergonzoni_iswap_2025,Muminov_2026}. Molecules are usually optically trapped during the logic operations, where the position spread inside the traps can not be ignored compared to the molecule-molecule separation $L$~\cite{ruttley_long-lived_2025,picard_entanglement_2025}, leading to an uncertainty of the DDI. Consequently, despite the remarkable advance of the cooling, trapping, and quantum state manipulation over ultracold polar molecules~\cite{Schindewolf_2022,Cornish_2024}, experimental demonstrations of quantum entanglement~\cite{Bao_2023,Holland_2023,ruttley_long-lived_2025} and two-qubit logic operations~\cite{picard_entanglement_2025} revealed that the uncertainty of DDI strongly limits the gate accuracy. Recently, Ref.~\cite{bergonzoni_iswap_2025} proposed an optimal control scheme that can yield a two-molecule iSWAP gate with an infidelity
below $2\times10^{-4}$ with 10\% variation of DDI.

Here, we introduce a way to realize a two-qubit molecular phase gate resilient to the uncertainty of DDI, and the gate is realized by two global microwave pulses when assisted by two single-qubit phase gates following each microwave pulse. Enabled by an exotic effective spin echo process, the gate has two features. First, variation of DDI barely influences the gate accuracy due to that DDI serves as a channel to enable adiabatic dynamics where fine tuning of DDI is no longer needed. Second, the entangling phase $\varphi$ in the gate is from the global microwave field, so that $\varphi$ can be of any desired value simply by tuning the relative phase between the microwave pulses. Such a class of tunable controlled phase gates can be used in quantum Fourier transform for notable quantum algorithms like, e.g., phase estimation, order-finding, and the Shor's algorithm~\cite{Nielsen2000}.

We introduce a theory to quantum mechanically study the influence of the molecular motion in the trap on the entangling gate. When neglecting the much weaker motion along the radial direction in typical tweezers, the molecular motion along the axial directions results in a coupling between the two molecules' harmonic motional modes and the DDI-coupled internal states. Quantum mechanically, this means that two motional modes are coupled to the internal states of the molecules via DDI. By a motional-mode separation technique, however, we find that though there are two modes involved in the molecular motion, only one is coupled to the internal state of the molecules. By this formalism, we find that the two-qubit phase gate can still acquire a gate fidelity beyond 0.9999 with typical molecular motion in traps of recent experiments~\cite{picard_entanglement_2025,ruttley_long-lived_2025}.

\section{ A controlled-phase gate by an effective spin echo}
For two molecules, labeled i and ii, we consider that for each molecule there are three different molecular states labelled $\lvert\uparrow\rangle$, $
\lvert\downarrow\rangle$, and $|e\rangle$, where the former two are qubit states, while the last is an ancillary state coupled with $\lvert\downarrow\rangle$ via a microwave field of a Hamiltonian
\begin{eqnarray}
\hat{H}_{\mu}&=&\hbar\frac{\Omega_{\mu}(t)}{2}\lvert e\rangle\langle\downarrow\rvert +\text{H.c.}, \label{Omega-mu}
\end{eqnarray}
which is in a rotating frame after performing the dipole and rotating-wave approximations. Since the microwave field in use has a wavelength on the order of decimeter while the molecular separation is on the order of micrometer, the same microwave field covers both molecules with equal strength and phase. In the Hilbert space spanned by the states of the two molecules, the microwave driving for the two molecules induces a Hamiltonian $\hat{H}_{\mu}\otimes\hat{I}+\hat{I}\otimes\hat{H}_{\mu}$ with $\hat{I}$ the identity operator. To induce entanglement, we consider a condition without applying static electric field and that a two-molecular state in a superposition of $\lvert\uparrow,e\rangle$ of $\lvert e,\uparrow\rangle$ experiences a DDI
$\hat{V} = J \left(\lvert \uparrow,e\rangle\langle e,\uparrow\rvert+ \lvert e,\uparrow\rangle\langle \uparrow,e\rvert\right)$~\cite{Ni_2018}. The absence of DDI in $\lvert \downarrow\rangle$ can be realized by selection rule or that DDI can not change nuclear spin states~\cite{Tscherbul_2023,picard_entanglement_2025}.

In the context of quantum logic gates, it is useful to study the gate map by examining the time dynamics for the energy eigenstates of the gate~\cite{ShiLu2024,Lu_2026}, which are $\lvert \uparrow,\uparrow\rangle,\lvert \uparrow,\downarrow\rangle,\lvert \downarrow,\uparrow\rangle,\lvert \downarrow,\downarrow\rangle$ here. The microwave field only couples $|e\rangle$ and $\lvert\downarrow\rangle$, so the state $\lvert \uparrow,\uparrow\rangle$ stays intact in the rotating frame. Both $\lvert \uparrow,\downarrow\rangle$ and $\lvert \downarrow,\uparrow\rangle$ are coupled by the microwave field, with dynamics that can be conveniently described by $\hat{H}=\hat{H}_++ \hat{H}_-$, with
\begin{eqnarray}
\hat{H}\pm &=&\left[\hbar\frac{\Omega_{\mu}(t)}{2}\lvert \mathbb{D}_\pm\rangle\langle\mathbb{B}_\pm\rvert +\text{H.c.}\right]\pm J\lvert \mathbb{D}_\pm\rangle\langle\mathbb{D}_\pm \rvert  ,\label{H_02}
\end{eqnarray}
where $\mathbb{D}$ and $\mathbb{B}$ are DDI-coupled Bell states and qubit-space Bell states, respectively, given by
\begin{eqnarray}
\lvert \mathbb{D}_\pm\rangle &=& \frac{1}{\sqrt{2}}\left( \lvert \uparrow,e\rangle\pm  \lvert e,\uparrow \rangle \right),\nonumber\\
\lvert \mathbb{B}_\pm\rangle &=& \frac{1}{\sqrt{2}}\left( \lvert \uparrow,\downarrow\rangle\pm  \lvert \downarrow,\uparrow \rangle \right).\label{DB-define}
\end{eqnarray}
Here, the use of Bell-state basis does not mean that the gate works in Bell-state basis. It is used only for showing the physical picture of the gate mechanism. Note that the numerical results show later are all from the original Hamiltonian without use of any Bell-state basis. $\hat{H}_\alpha$ can also be written as $\hat{H}_\alpha = \sum_{\eta =\pm }\xi_\eta^{(\alpha)} \lvert v_\eta^{(\alpha)}\rangle\langle v_\eta^{(\alpha)}\rvert $, where
\begin{eqnarray}
 \xi_\eta^{(\alpha)}&=&\frac{1}{2}\left(\alpha J +\eta \hbar\overline{\Omega}_\mu\right),\nonumber\\
 \lvert v_\eta^{(\alpha)}\rangle &=& \mathscr{N}_\eta^{(\alpha)}\left[ \xi_\eta^{(\alpha)}\lvert \mathbb{D}_\alpha \rangle +   \hbar\frac{\Omega_{\mu}^\ast(t) }{2} \lvert \mathbb{B}_\alpha \rangle   \right],\label{diag01}
\end{eqnarray}
where $\overline{\Omega}_\mu =\sqrt{J^2/\hbar^2 + |\Omega_\mu|^2}$ and $\mathscr{N}_\eta^{(\alpha)}=\left[ |\xi_\eta^{(\alpha)}|^2+     \hbar^2|\Omega_{\mu}(t)|^2/4 \right]^{-1/2} $ .

In a classical picture, $J$ is fluctuating or uncertain due to the molecular motion in the optical trap, so that any control scheme by a microwave driving will be subjected to a control error. To remove this error, we introduce an adiabatic spin echo sequence as follows. When $\Omega_\mu=0$ at $t=0$ and $J<0$, we have $
 \lvert \mathbb{B}_\pm \rangle = \lvert v_{\pm}^{(\pm)}\rangle$, and when we smoothly ramp up the microwave Rabi frequency to a maximum $\Omega$, and then back to zero again, we have
\begin{eqnarray}
 \lvert \mathbb{B}_\pm \rangle &\rightarrow& e^{\mp i\varphi}\lvert \mathbb{B}_\pm \rangle,\label{ada-002}\\
 \varphi &=& \int_0^T  \xi_{+}^{(+)}  dt/\hbar.\nonumber
\end{eqnarray}

After the first $\pi$ pulse, a phase gate that imprints a $\pi$ phase to the $\lvert \downarrow\rangle$ state of the molecule ii is applied. This operation needs single-site addressing, which can be via dynamic ramping of tweezer depths combined with microwave shelving~\cite{Picard_2024,picard_entanglement_2025}, or by laser fields to shift off non-targeted molecules out of resonance~\cite{Mortlock_2025}. The phase change on molecule ii results in
\begin{eqnarray}
 \lvert \mathbb{B}_\pm\rangle & \rightarrow& \lvert \mathbb{B}_\mp\rangle ,\label{ada-003}
\end{eqnarray}
where nothing occurs to the input state $\lvert \downarrow,\downarrow\rangle $ because it is $-\lvert e,e\rangle$ due to the already-applied first microwave pulse.

Then, a second $\pi$ pulse of microwave field with Rabi frequency $e^{-i(\pi+\varphi)/2}\Omega_\mu(t)$ is sent to the molecules. According to the adiabatic picture, we again have the state evolution as in Eq.~(\ref{ada-002}). Note that the phase shift to the microwave field does not alter the adiabatic picture. The only effect of this phase change is that for the input state $\lvert \downarrow,\downarrow\rangle $ which is $-\lvert e,e\rangle$ at the beginning of the second microwave pulse, we have an extra phase shift $\varphi+\pi$ when it returns back to the ground state. Then, the state evolution from the first microwave field to the end of the second microwave field is
\begin{eqnarray}
 \lvert \mathbb{B}_\pm \rangle &\rightarrow& \lvert \mathbb{B}_\mp \rangle,\nonumber\\
 \lvert \downarrow,\downarrow\rangle  &\rightarrow&-e^{i\varphi}\lvert \downarrow,\downarrow\rangle.
\end{eqnarray}
Finally, we use the phase shift gate as used for Eq.~(\ref{ada-003}), so that the reverse of Eq.~(\ref{ada-003}) occurs. The phase shift $\pi$ to the $\lvert \downarrow\rangle$ state of molecule ii also imprints a $\pi$ phase to $\lvert \downarrow,\downarrow\rangle$. As a consequence, the net effect for the two global microwave pulse and two single-qubit phase gates is $\lvert \downarrow,\downarrow\rangle\mapsto e^{i\varphi}\lvert \downarrow,\downarrow\rangle$ while all the other state components remain unchanged. For $\varphi=\pi$, the gate is the CZ gate which can be transformed to a controlled-NOT via single-qubit rotations~\cite{Shi2021qst}.

\begin{figure}
\includegraphics[width=3.40in]
{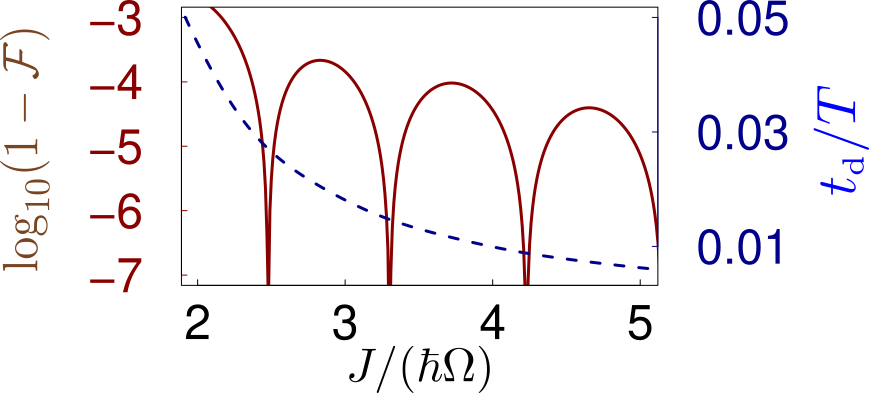}
\caption{The solid curve shows log$_{10}$(infidelity) of the CZ gate, i.e., a two-qubit gate with a controlled phase $\varphi=\pi$, simulated with the pulse in Eq.~(\ref{omega-time}). Here, the oscillation of the infidelity when varying $J$ arises from the detuned nature of the Rabi cycles, which is a common feature in detuned Rabi oscillation~\cite{Shi2018prapp2}. The dashed curve shows the total time being in the DDI coupled states $\lvert e,\uparrow\rangle$ and $\lvert \uparrow,e\rangle$ for the initial state $\lvert \uparrow,\downarrow\rangle$; The initial state $\lvert \downarrow,\uparrow\rangle$ has the same $t_{\text{\tiny d}}$. When averaging over the four input eigenstates, the average time to state in a DDI-coupled state is $t_{\text{\tiny d}}/2$.   \label{figure-uni-error} }
\end{figure}

 To examine how much can the above picture tolerate with a moderate $J/\Omega$, we take, as an example,
\begin{eqnarray}
  \Omega_\mu(t) &=&\Omega \left[e^{-\frac{(t-T/2)^2}{2t_{\text{w}}^2}}-e^{-\frac{T^2}{8t_{\text{w}}^2}}\right],~\text{when }t\leq T,\nonumber\\
  \Omega_\mu(t)&=& i\Omega_\mu(t-T),~\text{when }t\in (T,~2T],
  \label{omega-time}
\end{eqnarray}
and set $t_{\text{w}}=0.234T$ at which $\Omega$ is smallest, equal to about 6.75$/T$, while simultaneously satisfies $\int_0^T \Omega_\mu dt=\pi$. We have simulated the time dynamics with the pulse of Eq.~(\ref{omega-time}) and a Hamiltonian in the basis of $\{\lvert \uparrow,\uparrow\rangle,\lvert \uparrow,\downarrow\rangle,\cdots\}$, based on which we calculate the infidelity of the gate with the definition of fidelity from Ref.~\cite{Pedersen2007}. The results are shown in Fig.~\ref{figure-uni-error}, which also shows the DDI-superposition time of the initial state $\lvert \uparrow,\downarrow\rangle$, i.e., time to stay in $\lvert e,\uparrow\rangle$ and $\lvert \uparrow,e\rangle$,
\begin{eqnarray}
  t_{\text{\tiny d}} &=&\sum_{\vert\eta\rangle \in\{\lvert e,\uparrow\rangle,\lvert \uparrow,e\rangle\}} \int \left|\langle \eta \lvert\mathcal{T} e^{-\frac{i}{\hbar}\int_0^\tau \hat{H}t}|\uparrow,\downarrow\rangle\right|^2,
\end{eqnarray}
which, when being small, indicates that $\lvert \uparrow,\downarrow\rangle$ is barely excited. If we suppose that the desired $J$ is $4\hbar\Omega$, with $\hbar$ the reduced Planck constant, and consider its variation up to a quarter of its desired value, then we can take an average of $J$ over $[3,~5]\hbar\Omega$, i.e., a 25\% variation of DDI, leading to an average gate fidelity 0.99996. This remarkable robustness is due to that the gate has little dependence on populating the state in a DDI coupled state as indicated by the smallness of $t_{\text{\tiny d}}$. Because there are four input eigenstates, the average time for an input state to stay in a DDI-coupled state is $t_{\text{\tiny d}}/2$.

\section{Influence from position spread in traps}
To capture the effect of the finite position spread of optically trapped molecules on the entangling gates, a quantum mechanical treatment is desirable. When we approximate the trap as harmonic~\cite{ruttley_long-lived_2025}, the position spread arises due to that the desired trap centers for the two molecules are $L$, while there is finite extension of the positions of the molecules. The trap frequencies along the axial and radial directions differ by one order of magnitude in usual experimental setups~\cite{ruttley_long-lived_2025}. As a result, the position spread of the trapped molecules is mainly along the axial direction. So, the much smaller position spread along the radial direction can enable the neglect of the motion along the radial direction as in Ref.~\cite{picard_entanglement_2025}. When the angular frequency is $\omega$ along the axial direction, the harmonic oscillator length is $
\ell =  \sqrt{\hbar/(m\omega)}$, where $m$ is the mass of the molecule. To have strong DDI, $L$ is usually quite small, around 2~$\mu$m in recent experiments~\cite{ruttley_long-lived_2025,picard_entanglement_2025}. As a result, the position spread should not be neglected, with $\ell/L$ about $0.05$ in tweezer-trapped $X^1\Sigma^+$ NaCs molecules of Ref.~\cite{picard_entanglement_2025}.

So, we can consider the centers for two traps at $(0,0,0)$ and $(0, L, 0)$, with the axial direction along $\mathbf{x}$. Along the axial direction, the Hamiltonian for the two trapped molecules is
\begin{eqnarray}
 \hat{H}_{\text{trap}} &=&  \sum_{\alpha =\text{i,ii}}\left[ \frac{p_{\alpha}^2}{2m} +\frac{m\omega^2x_{\alpha}^2}{2}\right] ,\label{H-0}
\end{eqnarray}
where $\alpha=$i, ii, and $p_\alpha=-i\hbar\partial/\partial x_\alpha$. To capture the coupling between the external motion and the internal rovibrational states, we define
\begin{eqnarray}
 x_{\pm}&=&\frac{1}{\sqrt{2}}(x_{\text{i}}\pm x_{\text{ii}}),~
 p_{\pm}= -i\hbar \frac{\partial}{\partial x_\pm}, \label{transform}
\end{eqnarray}
so that Eq.~(\ref{H-0}) becomes,
\begin{eqnarray}
 \hat{H}_{\text{trap}} &=&  \sum_{\alpha =\pm}\left[ \frac{p_{\alpha}^2}{2m} +\frac{m\omega^2x_{\alpha}^2}{2}\right] ,\label{H-1}
\end{eqnarray}
which means that the state of the external two-molecule vibration in the two-trap system is a product state of the two motional modes labeled $\pm$. The wavefunctions of the ground and the $n$th-excited two-trap motional state of the molecules are $|0_{\alpha}\rangle\equiv\lvert\text{vac}_\alpha\rangle$ and $|n_{\alpha}\rangle$, with $n=1,~2,~\cdots$. Here, $|n_{\alpha}\rangle = \frac{1}{\sqrt{2^nn!\ell\sqrt{\pi}}} e^{-\frac{x_\alpha^2}{2\ell^2}}h_n(x/\ell)$, and the Hermitian polynomials $h_n(\xi)$ are given by $h_n(\xi)= (-1)^ne^{\xi^2}\frac{\partial^n}{\partial \xi^n}e^{-\xi^2}$
which is an eigenstate of eigenenergy $E_n= \frac{1}{2}\hbar \omega(2n+1)$. In the Hilbert space spanned by $|n_{\alpha}\rangle$, the motional Hamiltonian is
\begin{eqnarray}
 \hat{H}_{\text{trap}} &=& \frac{1}{2}\hbar \omega\sum_{\alpha=\pm}(2\hat{a}_\alpha^\dag \hat{a}_\alpha+1),
\end{eqnarray}
where $\hat{a}_\alpha^\dag$ and $\hat{a}_\alpha$ are the creation and annihilation operators for the motional states of the two-molecule motional mode $\alpha=$+ or -, which is related to $x_\alpha$ via $x_\alpha = \frac{\hat{a}_\alpha^\dag + \hat{a}_\alpha }{\sqrt{2}}\ell$. Then, the modes $\pm$ are related to the modes i and ii via $\hat{a}_{\pm}=\frac{1}{\sqrt{2}}(\hat{a}_{\text{i}} \pm\hat{a}_{\text{ii}}  )$
and similarly for the creation operator $\hat{a}_{\pm}^\dag$. So, if the two molecules are both in the motional ground states, the state in the basis of $\alpha=\pm$ is also the ground one.

\begin{figure}
\includegraphics[width=3.0in]
{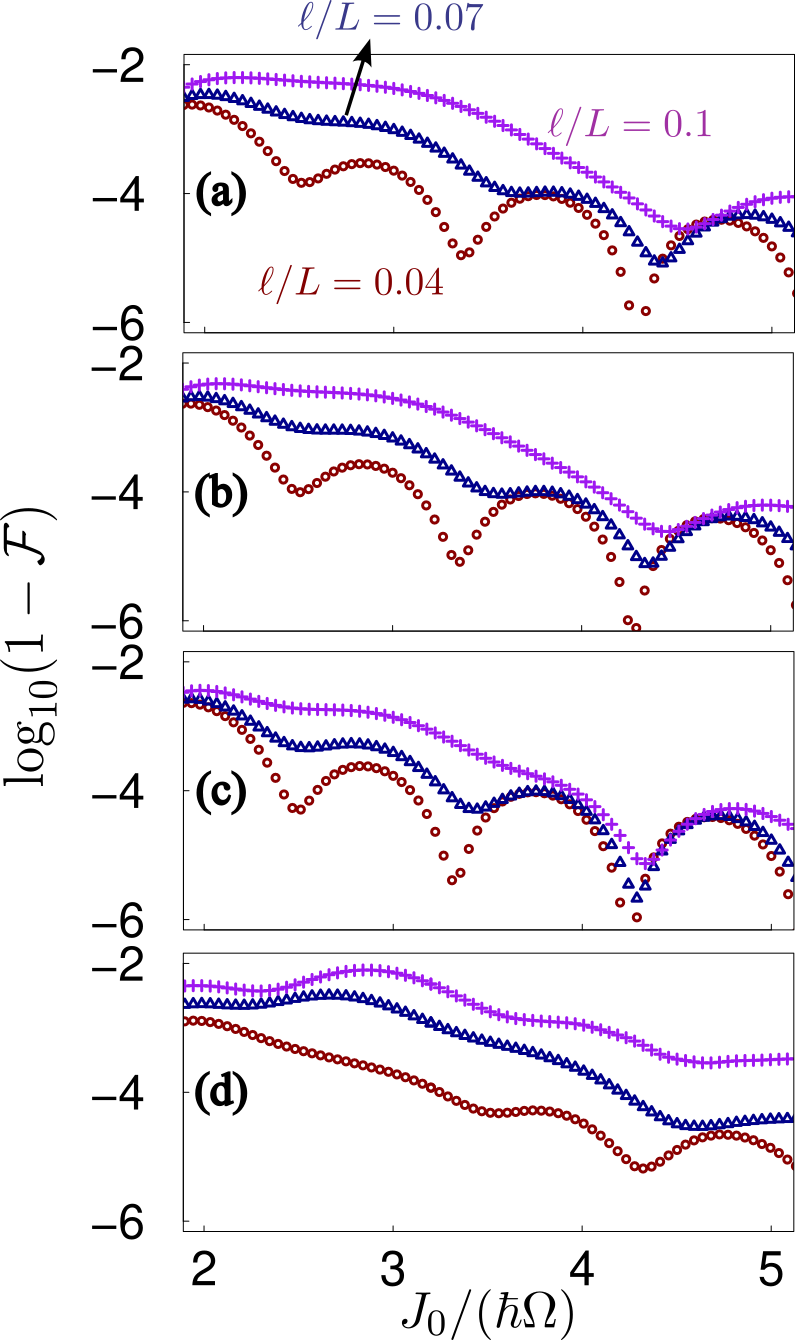}
\caption{Round, triangle, and cross symbols show log$_{10}$(infidelity) of the CZ gate with $J$ given in Eq.~(\ref{J-fluctuation4-0}), and $\ell/L=0.04,0.07$ and $0.1$, respectively. Here, $\ell$ is the harmonic oscillator length and $L$ the distance between the centers of the two tweezers. The initial states for the $\hat{a}_-$ mode of the motion are $\hat{a}_-^\dag\lvert \text{vac}_-\rangle, (1+\hat{a}_-^\dag)\lvert \text{vac}_-\rangle/\sqrt{2},$ and $\lvert \text{vac}_-\rangle$ in (a, b), and (c), respectively. The initial state in (d) is a thermal state with $\langle\hat{a}_-^\dag\hat{a}_-\rangle=2$. States up to 40 motional excitations of the $\hat{a}_-$ mode are included in the simulation via QuTip~\cite{Johansson_2012,Johansson_2013}, at which we find the simulation converged to a high level; we found no difference between results with 40 motional excitations and those with 100 motional excitations. Parameters are the same as those in Fig.~\ref{figure-uni-error}, except that here we need another parameter, i.e., the trap frequency, for which we set $\omega=\Omega$ for brevity because both $\Omega$ and $\omega$ can be in the kHz regime~\cite{ruttley_long-lived_2025,picard_entanglement_2025}. \label{figure-uni-error2} }
\end{figure}

The modes defined in Eq.~(\ref{transform}) are useful for studying the coupling between DDI and molecular motion in the traps. When the quantization axis is along $\mathbf{y}$, the coupling strength of the DDI can be written as
\begin{eqnarray}
J &\approx& J_0\left[\frac{3\ell^2}{L^2}\left(\hat{a}_-^{\dag}+  \hat{a}_- \right)^2-\frac{45\ell^4}{8L^4}\left(\hat{a}_-^{\dag}+  \hat{a}_- \right)^4-1\right],\nonumber\\ \label{J-fluctuation4-0}
\end{eqnarray}
which shows that the DDI only couples with the $\hat{a}_-$ mode. In other words, if the two-trap motional state is in a state that the $\hat{a}_+$ mode is highly excited, it still has no influence on the DDI because only the $\hat{a}_-$ mode is involved.

For typical recent experiments, the molecules in the traps are in the ground and the first excited states, with little chance to be in a higher state~\cite{ruttley_long-lived_2025}. When converting from the original modes to the $\hat{a}_\pm$ mode, one can see that the frequencies for the four modes are equal. In other words, if we assume that the molecules are cooled to states with no more than one motional excitation, then the worst case in the relevant $\hat{a}_-$ mode is that we have one vibration quantum in it.

The major influence from the motional states is that states with more motional excitations contribute more to the change of motional states, which is detrimental because it results in undesired entanglement between the external motion and the internal states. For analyzing the gate fidelity, we take both pure motional states and thermal states. Figure~\ref{figure-uni-error2}(a,~b) and (c) show high fidelity with three representative pure motional states $\hat{a}_-^\dag\lvert \text{vac}_-\rangle\otimes\lvert\text{A}_+\rangle, [(1+\hat{a}_-^\dag)\lvert \text{vac}_-\rangle/\sqrt{2}]\otimes\lvert\text{A}_+\rangle,$ and $\lvert \text{vac}_-\rangle\otimes\lvert\text{A}_+\rangle$, respectively, with $\lvert\text{A}_+\rangle$ an arbitrary state for the $\hat{a}_+$ mode of the motion. Figure~\ref{figure-uni-error2}(d) shows results when the initial motion is a thermal state with $\langle\hat{a}_-^\dag\hat{a}_-\rangle=2$. One can see that for $J/(\hbar\Omega)$ around 4, the infidelity is on the order of $10^{-4}$ in Figs.~\ref{figure-uni-error2}(a,~b) and (c) even with $\ell/L$ up to $0.1$, and Fig.~\ref{figure-uni-error2}(d) shows that $\mathcal{F}$ is over 0.9999 with $\ell/L=0.04$, which is near the condition of Ref.~\cite{picard_entanglement_2025}.

Though Fig.~\ref{figure-uni-error2} shows that the gate can be extremely accurate even with a large $\ell/L$, it is useful to suppose that there can be certain uncertainty of the trap centers, leading to variation of $J_0$ with respect to a desired value. One can average the fidelity by $J_0$ over, e.g., $[3,~5]\hbar\Omega$, i.e., a 25\% change, with results shown in Table~\ref{table-0}. For the experiments in Ref.~\cite{picard_entanglement_2025}, $\ell\approx0.1~\mu$m and $L\in[1.79,~2.5]~\mu$m, yielding $\ell/L\in(0.04,~0.06)$. According to Table~\ref{table-0}, the influence of the coupling to motional states with $\ell/L\in(0.04,~0.06)$ is quite insignificant.

 The example shown in Figs.~\ref{figure-uni-error},~\ref{figure-uni-error2} and Table~\ref{table-0} reveals that if $\ell/L=0.04$ near the condition of Ref.~\cite{picard_entanglement_2025}, we have $\mathcal{F}>0.9999$ even if the molecules are in a thermal state of the $\hat{a}_-$ mode with two motional quanta. In Ref.~\cite{ruttley_long-lived_2025}, about 58\% of molecules occupy the motional ground state, and most of the motional excited molecules have just one motional quantum. In practice, it is not trivial to have the molecules in a pure motional state, so that it is possible that the results in Figs.~\ref{figure-uni-error2}(d) is more practical. This means that our gate can obtain a gate fidelity over 0.9999 with currently available experimental conditions.

\begin{table}[ht]
  \centering
  \begin{tabular}{|c|c|c|c|}
    \hline
\diagbox{$\hat{a}_-$ state}{$\ell/L$} & $0.04$ & $0.07$ &  $0.1$ \\
     \hline
$\hat{a}_-^\dag\lvert \text{vac}_-\rangle$ &  0.99995      & 0.99983     & 0.99906\\\hline$(1+\hat{a}_-^\dag)\lvert \text{vac}_-\rangle/\sqrt{2}$ & 0.99995  &  0.99988 & 0.99940\\
   \hline $\lvert\text{vac}_-\rangle$ &0.99996 & 0.99993  &0.99974\\
   \hline \text{Thermal, }$\langle\hat{a}_-^\dag\hat{a}_-\rangle=2$ &0.99995 & 0.99957  &0.99830\\
   \hline
  \end{tabular}
  \caption{Average gate fidelity with $J_0$ equally sampled over $[3,~5]\hbar\Omega$ for three representative pure initial states of the $\hat{a}_-$ motional mode, and a thermal initial state with two motional quanta in the $\hat{a}_-$ mode. Parameters used are the same to those in Fig.~\ref{figure-uni-error2} \label{table-0}  }
  \end{table}

\section{Discussion and Conclusions}
The gate here is analogous to dark-state gates in Rydberg atoms, for a review, see, e.g., Ref.~\cite{Shi2021qst}. One difference from Rydberg atoms is that the dark-state gates with Rydberg atoms need populating multiple Rydberg states which have lifetimes of order of 0.1~ms, but the DDI occurs between molecules in low-lying states. Here, $\lvert \uparrow\rangle$ and $\lvert\downarrow\rangle$ can be in the ground rovibrational manifold and $\lvert e\rangle$ in the first excited manifold, all possessing long lifetimes. A particular strength of polar molecules is that it is possible to trap multiple low-lying states in the same trap with second scale coherence, while simultaneously enabling coherent state transitions between the states~\cite{Hepworth_2025}. Therefore high-fidelity controlled phase gates with polar molecules are realizable.

In summary, we introduce a controlled phase gate with two close polar molecules implemented by two consecutive $\pi$ pulses of microwave fields when assisted by two single-qubit phase gates. The gate acquires a fully tunable phase determined by the phase change from the first to the second microwave pulses. We further introduce a quantum mechanical treatment which can help to analyze the molecular motion in the trap during the gate operation, and find that a gate fidelity over 0.9999 is possible with recent experimental setups.

\section*{acknowledgments}
We acknowledge the National Natural Science Foundation of China under Grant Nos. 12074300 and 12547103, and the Innovation Program for Quantum Science and Technology 2021ZD0302100 for support, and thank all the speakers in the Haikou winter school ``Fundamentals and Frontiers of ultracold atoms and molecules'', and Xin-Yu Luo, Xiangchuan Yan, and Peter Schmelcher for discussions.

%

\end{document}